# Sistem Informasi Eksekutif Berbasis Web Pada Fakultas Pertanian Universitas Muhammadiyah Palembang

Fauzan Aziz[1], Leon Andretti Abdillah[2], Novri Hadinata[3]
[1,2,3] Program Studi Sistem Informasi, Fakultas Ilmu Komputer, Universitas Bina Darma
Palembang, Indonesia
[1]fauzan.binas@gmail.com, [2]leon.abdillah@yahoo.com

**Abstract.** Information technology (IT) is able to fulfill one of the main needs of an organization, such as how the executive know and manage the performance of the organization he leads, including the human resources (HR). Faculty of Agriculture, University of Muhammadiyah Palembang (UMP) has had personnel information system which is used to manage HR data both employees and lecturers. However, the information system is to support the operational activities only. Therefore it is necessary to build an executive information system (SIE) in the faculty is the dean. By using the SIE, the dean can easily access summary data visualization, namely the appearance of information in the form of graphs making it easier for executives to make decisions.

## 1 Pendahuluan

Sistem informasi (SI) telah banyak berkembang dengan pesat dan membantu manusia dalam menyelesaikan perkerjaannya. Salah satunya adalah sistem informasi eksekutif (SIE). SIE adalah sistem informasi yang digunakan untuk para eksekutif dalam mengambil keputusan strategis. Ketersediaan akses informasi strategis secara langsung bagi para eksekutif sangat diperlukan karena para eksekutif memiliki peran utama sebagai pengambil keputusan strategis dan taktis. SIE merupakan sisem informasi yang menyediakan fasilitas yang fleksibel bagi manager dan eksekutif dalam mengakses informasi eksternal dan internal yang berguna untuk mengidentifikasi masalah atau mengenali peluang [1].

Aplikasi berbasis *web* adalah sebuah program yang disimpan dan dikirim melalui *internet* dan diakses melalui antarmuka *browser*. Dengan begitu aplikasi yang dibangun dapat diakses kapan pun dan dimana pun selagi dapat terhubung dengan *internet*. Sehingga tepat apabila SIE berbasis *web* dikembangkan, karena dapat lebih mempermudah eksekutif dalam mengakses SIE tersebut.

Fakultas Pertanian Universitas Muhammadiyah Palembang (UMP) telah mempunyai sistem informasi kepegawaian (SIK) yang digunakan untuk mengolah data sumber daya manusia (SDM) baik itu karyawan maupun dosen. Namun, sistem informasi tersebut lebih menunjang kegiatan operasional sehingga belum menyajikan informasi atau laporan khusus bagi eksekutif khususnya dekan Fakultas Pertanian.

Berdasarkan observasi penyajian informasi kepada eksekutif khususnya dekan lebih banyak disajikan dalam bentuk dokumen ataupun lisan. Dengan demikian proses mendapatkan informasi penting yang dapat menunjang pengambilan keputusan strategis oleh eksekutif memerlukan waktu yang tidak cepat. Idealnya, para eksekutif dapat mengakses informasi strategis secara *real-time* dan mandiri [2].

Berdasarkan uraian di atas maka sangat tepat apabila UMP terutama pada Fakultas Pertanian menerapkan suatu strategi yang lebih baik dengan cara membangun SIE Berbasis Web. Sistem ini akan mendukung eksekutif dalam pengambilan keputusan serta merencanakan strategi selanjutnya karena informasi yang disajikan nantinya lebih cepat, mudah, ringkas, interaktif dan keluar dari pendekatan tradisional kebutuhan dimana laporan diberikan dalam bentuk visual berupa grafik.

Setelah mengulas sejumlah penelitian terdahulu, maka SIE yang akan dibangun akan : 1) diperuntukkan bagi *top-level management* atau eksekutif untuk mengambil keputusan [3], 2) sesuai dengan kebutuhan [4] dalam mengontrol dan mengawasi kinerja perusahaan yang dipimpinnya secara ringkas, terintegrasi, mudah dipahami, dan dalam berbagai tingkatan rincian [5], 3) ada visualisasi data [3] yang ditampilkan dalam bentuk *dashboard* [6] yang terdiri atas grafik ber-kapabilitas *drill-down* [2, 3].

## 2   Metode Penelitian

Metode yang digunakan peneliti untuk mengembangkan SIE Berbasis Web pada Fakultas Pertanian UMP adalah siklus hidup sistem informasi eksekutif (*EIS Lifecycle*) [7] yang terdiri dari: 1) justikasi, 2) perencanaan, 3) analisis bisnis, 4) desain sistem, 5) kontruksi, dan 6) sistem *deployment*.

### 2.1   Justifikasi (*Justification*)

Dalam tahap ini dilakukan pengidentifikasian *business case assesment* pada bidang SDM dengan teknik wawancara kepada pihak terkait. Sejumlah kebutuhan dan peluang diantaranya sebagai berikut: 1) Penyajian informasi kepada eksekutif khususnya dekan lebih banyak disajikan dalam bentuk dokumen ataupun lisan, 2) Dekan sudah seharusnya mendapatkan informasi terkini yang mereka butuhkan dalam pengambilan keputusan yang menyangkut bidang kepegawaian secara *up to date*, dan 3) SIE dapat menjadi alternatif dalam memecahkan permasalahan karena kemampuannya untuk memberikan informasi bagi para eksekutif secara ringkas, terintegrasi, mudah dipahami, dan dalam berbagai tingkatan rincian.

## 2.2 Perencanaan (*Planning*)

Pada tahap perencanaan ini dilakukan identifikasi infrastuktur organisasi dengan teknik observasi. Tahapan perencanaan menjelaskan *enterprise infrastructure evalution* dan *project planning*.

Kegiatan pada tahapan ini yaitu melakukan identifikasi infrastruktur organisasi pada Fakultas Pertanian UMP. Infrastruktur sat ini sudah menggunakan aplikasi komputerisasi, serta ada jaringan LAN dan internet sebagai pendukung SIE.

Tahap *project planning* yang akan dilakukan pada awal bulan April sampai dengan Agustus 2015. Fakultas Pertanian UMP berlokasi di Jl. Jendral Ahmad Yani, 13 Ulu Palembang.

## 2.3 Analisis Bisnis (*Business Analysis*)

Analisis bisnis yaitu mengetahui kebutuhan bisnis yang ada pada bidang SDM Fakultas Pertanian UMP. Kebutuhan sistem dari 2 (dua) aktor: 1) Dekan dapat melihat laporan dari data pegawai yang meliputi data dosen berdasarkan umur, jenjang S3 dosen, dosen pns dan belum pns, pendidikan dosen. Data karyawan berdasarkan status karyawan, masa kerja karyawan, dan umur karyawan, dan 2) Bagian SDM: a) melakukan pendataan pegawai meliputi penambahan dan perubahan data baik data karyawan dan data dosen, dan b) dapat melihat hasil laporan.

## 2.4 Desain Sistem (*System Design*)

Dalam tahap ini dilakukan desain terhadap kebutuhan informasi bagi seorang eksekutif. Tahapan *desain system* terdiri dari: 1) *data design*, 2) *designing ETL process* (*extract/transform/load*), dan 3) *metadata repository design*.

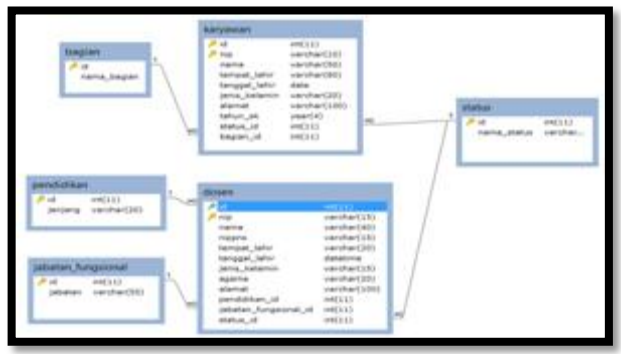

**Gambar 1.** *Desain data*

Pada tahap *data design*, pembuatan model *logical* dilakukan untuk mendapatkan model tabel yang akan digunakan untuk menyimpan data dari sumber data yang sebelumnya sudah didiperoleh. Adapun alat yang digunakan untuk menyimpan data kebasis data SIE adalah *database Mysql* (gambar 1).

Dalam rancangan proses ETL dilakukan dengan mengambil data yang ada pada file Microsoft Excel. Dari data ini akan dilakukan proses ekstrak ke dalam *database* yang dibangun dengan menggunakan *database MySQL*. Selanjutnya *database* ini yang akan menjadi *data warehouse* (gambar 2).

Pada tahap *Metadata repository design* dilakukan perancangan *Metadata repository* dari sumber data diunggah ke basis data SIE pada proses *Extract Transform Load* (ETL). Data yang telah mengalami pembersihan dan pengintegrasian kemudian digunakan sebagai gudang data SIE.

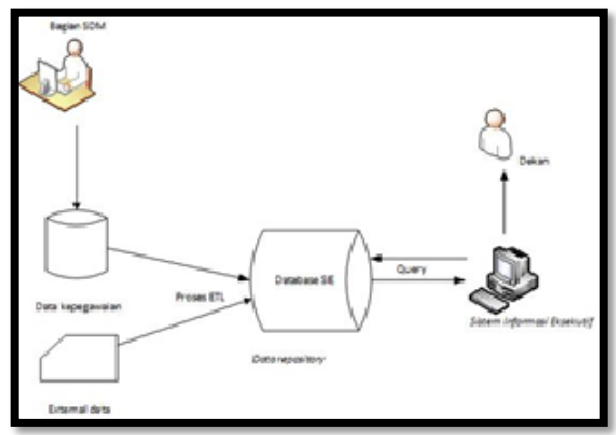

**Gambar 2.** *Design ETL Process*

## 2.5 Konstruksi (*Construction*)

Pada tahap ini menjelaskan tentang rekayasa sistem yang akan dibangun. Adapun tahap kontruksi yaitu: 1) *ETL development*, 2) *Application development*, 3) *Data Mining*, dan 4) *Developing metadata repository*.

Tahap *ETL development*. Setelah designing ETL process dilakuan maka didapatlah data-data yang sudah mengalami pembersihan dan siap digunakan didalam SIE nantinya. Yang terdiri dari data karyawan, data dosen, data bagian, data pendidikan, data status dan data jabatan fungsional. alat bantu yang digunakan peneliti adalah *ESF Database Migration Toolkit*.

Tahap *Application development*. Pada tahap ini dilakukan pembagian hak akses yang terdiri dari dua bagian yaitu bagian SDM dan dekan.

Tahap *Data mining*. Proses yang dilakukan dalam tahapan *data mining* mengunakan metode seleksi. Seleksi data dilakukan sesuai bentuk yang diinginkan. Proses seleksi ini menggunakan *Structure Query Language (SQL)* pada *MySql*. Contoh: Seleksi untuk menampilkan data dosen berdasarkan umur

```
SELECT nama, YEAR(CURDATE())-YEAR(`TgLahir`) AS umur
FROM dosen
```

Tahap *Development metadata repository* membuat rincian tabel dari *metadata repository design* yang sebelumnya sudah dilakukan terdiri dari tabel dosen, karyawan, status, bagian, jabatan fungsional, pendidikan dan *users*.

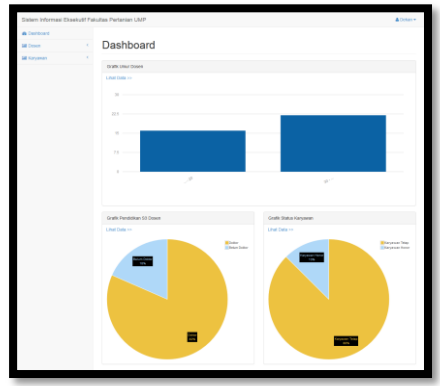

**Gambar 3.** *Dashboard*

## 3  Hasil dan Pembahasan

Berdasarkan hasil penelitian yang dilakukan pada Fakultas Pertanian UMP, maka didapatkan hasil akhir sebuah sistem yaitu SIE berbasis *web* bidang SDM Fakultas Pertanian UMP.

### 3.1  *Dashboard*

Halaman ini merupakan halaman yang menampilkan informasi berupa grafik yang terdiri dari : 1) grafik umur dosen, 2) grafik jenjang pendidikan dosen, dan 3) grafik status karyawan (gambar 3).

### 3.2  Informasi Mengenai Dosen

Informasi mengenai dosen berupa informasi rekapitulasi dosen berdasarkan : 1) kategori usia dosen, 2) jenjang pendidikan dosen, 3) jabatan fungsional akademik dosen, dan 4) status dosen. Contoh informasi dosen dapat dilihat pada gambar 4.

### 3.3 Informasi Mengenai Karyawan

Informasi mengenai karyawan berupa informasi rekapitulasi karyawan berdasarkan: 1) Status karyawan, 2) Masa keja karyawan, 3) Kateogori usia, 4) Data bagian, dll. Contoh informasi karyawan dapat dilihat pada gambar 5.

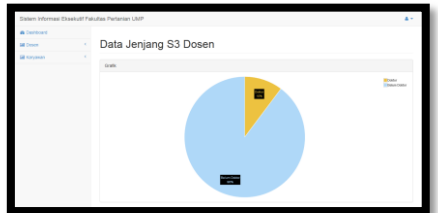 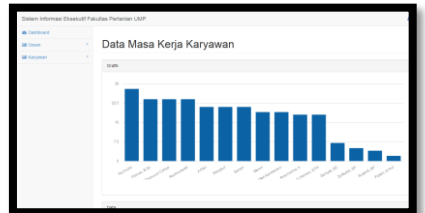

**Gambar 4.** Jenjang Pendidikan Dosen    **Gambar 5.** Masa Kerja Karyawan

## 4 Kesimpulan

Berdasarkan uraian yang telah dikemukakan pada hasil dan pembahasan sebelumnya, maka dapat ditarik beberapa kesimpulan dalam mencapai tujuan yang diinginkan. Adapun kesimpulan yang dapat diambil adalah sebagai berikut :
1. SIE ini dapat memberikan informasi ringkas dan mudah dipahami dekan Fakultas Pertanian UMP dengan tampilan *dashboard* berupa grafik.
2. Mempermudah dekan dalam mengambil keputusan jangka panjang. Sehingga dapat mempengaruhi kinerja dari bagian kepegawaian tersebut.

## Daftar Pustaka